# Applications of Reconfigurable Intelligent Surface in Smart High Speed Train Communications

ZHAO Yajun[1,2]

ZHANG Jiayi[3]

AI Bo[3]

(1. State Key Laboratory of Mobile Network and Mobile Multimedia Technology, Shenzhen 518057, China;

2. ZTE Corporation, Shenzhen 518057, China;

3. State Key Lab of Rail Traffic Control & Safety, Beijing Jiaotong University, Beijing 100044, China)

**Abstract:** as one of the most potential 5g adv and 6G key technologies, Reconfigurable Intelligent Surface (RIS) has the characteristics of low cost, low complexity and easy deployment, which provides a new opportunity for the development of intelligent high-speed rail communication. This paper introduces the typical applications of RIS assisted intelligent high-speed rail communication, including suppressing Doppler shift effect, solving the problem of frequent switching, overcoming the problem of high penetration loss and supporting high-precision train positioning. The key technologies of RIS assisted intelligent high-speed rail communication are discussed, including channel measurement and modeling, channel estimation and feedback, beamforming, network architecture and deployment. It is believed that the combination of the new infrastructure of intelligent high-speed rail and the new electromagnetic infrastructure constructed by RIS will bring broad technical and industrial prospects to intelligent high-speed rail in the future.

**Abstract:**
Reconfigurable intelligent surface (RIS) is one of the most promising technologies for 5G-Adv and 6G. It has the characteristics of low cost, low complexity, and easy deployment, which provides a new opportunity to develop intelligent high-speed rail communications. The typical applications of RIS-assisted smart high-speed railway communications are introduced in detail, including suppressing the Doppler shift effect, solving frequent handoff problems, overcoming high penetration loss problems, and supporting high-precision train positioning. The key technologies of RIS-assisted smart HST communications are discussed in-depth, including channel measurement and modeling, channel estimation and feedback, beamforming, network architecture, and network deployment. It is believed that the combination of the new intelligent high-speed railway infrastructure and the new electromagnetic infrastructure built by RIS will bring broad technological and industrial prospects to the intelligent high-speed railway in the future.

In the past 10 years, with the rapid development of high-speed railway (hereinafter referred to as high-speed railway) and the progress and integration of mobile communication technology and artificial intelligence (AI) technology, high-speed railway has begun to evolve from informatization to intelligence. 5g network naturally supports the interconnection of all things, so its large-scale commercial use will accelerate the development of high-speed rail intelligence. Although 5g technology can support the "high speed" and "intelligence" of intelligent high-speed rail communication, its engineering implementation is extremely challenging.

As a new technology, Reconfigurable Intelligent Surface (RIS) has attracted the attention of the industry as soon as it appeared. In the past two years, RIS has developed rapidly in academic research and industrial promotion, and is considered to be one of the key candidate technologies for the future mobile communication network [1-2]. RIS is usually composed of a large number of carefully designed electromagnetic units. By applying control signals to the adjustable elements on the electromagnetic units, the electromagnetic properties of these electromagnetic units are dynamically controlled, and then the space electromagnetic wave is intelligently regulated in a programmable way to form an electromagnetic field with controllable amplitude, phase, polarization and frequency. As a two-dimensional implementation of metamaterials, RIS naturally has the characteristics of low cost, low complexity and easy deployment, which can better meet the challenges of intelligent high-speed rail communication scenarios.

There are many researches on the wireless communication application of RIS technology in classic scenarios, but there are few literatures systematically discussing its application in high-speed rail communication [3], and only a few literatures involve the single point technology in this scenario. For example, reference [3] provides the idea of using RIS to reduce Doppler effect.

## 1. Requirements and Challenges of Smart high-speed rail Mobile Communications

The main research contents of intelligent high-speed rail mobile communication technology include broadband mobile communication, on-board wireless communication, intelligent dispatching communication, vehicle ground / vehicle vehicle communication technology, etc., which are used to support four types of typical services under the scenario of intelligent high-speed rail mobile communication, including train control and operation related services, train integrated service business, railway Internet of things business and passenger on-board mobile broadband access business [5].

Compared with the classic communication scenario, the wireless communication of intelligent high-speed rail is currently facing many severe challenges due to its wireless transmission environment and business characteristics, mainly including: (1) the serious Doppler frequency shift and frequent cell handover problems caused by the ultra-high speed movement of high-speed rail; (2) The high penetration loss of high-speed railway carriage makes it difficult to improve the signal coverage in the carriage; (3) It is necessary to make full use of more frequency bands (including those below 6 GHz and millimeter wave bands) to better meet the future requirements of intelligent high-speed wireless rail communication, so wireless networks are required to have the ability to support multiple frequency bands; (4) The wireless communication network of intelligent high-speed railway can be covered by private network or public network, so it faces the problem of complex coexistence of different systems with the surrounding networks along the railway; (5) Other problems, such as channel estimation and feedback brought by high-speed movement, high-precision positioning and environmental awareness of trains, and the coexistence of multiple service types in intelligent high-speed rail communication.

In order to solve the above problems, the traditional high-speed railway wireless communication system mainly adopts the methods of optimizing the transceiver algorithm and the network deployment of the transceiver. For example, the distributed large-scale antenna technology, the transmitter Doppler estimation and pre compensation technology and the handover process optimization technology are adopted. These traditional methods have high system complexity, difficult network deployment and optimization, and high implementation cost.

In addition, although the industry has a lot of research on AI to enhance the traditional wireless communication system, it mainly discusses the intelligence of the transmitter and receiver, and the wireless channel still needs to passively adapt to the natural propagation environment. The root of the challenge of intelligent high-speed rail wireless communication lies in its complex wireless channel environment. If we can control the wireless propagation environment artificially, we can fundamentally eliminate the unique complex channel environment of high-speed rail communication, which is obviously a better choice. On the basis of realizing the intelligentization of the wireless system's receiver and transmitter, further realize the intelligent control of the wireless channel, and can build an end-to-end intelligent wireless system that truly covers the transmitter, wireless channel and receiver.

To sum up, traditional high-speed rail wireless communication solutions can only passively adapt to the channel characteristics of high-speed rail. This is a great challenge to the complexity and cost of intelligent high-speed rail communication. The emergence of RIS allows people to regulate the wireless transmission environment and build an intelligent and controllable wireless transmission environment, which provides an opportunity to jointly optimize the transceiver and wireless transmission channel, and provides the possibility to further improve the system performance and reduce the complexity and cost.

## 2 Typical Application of RIS Enabled Intelligent High-speed Rail Wireless Communication

Smart high-speed rail wireless communication faces many challenges, but it also has significant regularity. For example, the wireless channel changes regularly along the railway line; The overall migration of the user terminal (UE) Group on the train has significant commonalities, including the group handover of the overall movement of the UE group, the overall migration of service connections and capacity requirements. Based on these characteristics, the wireless communication algorithm and network deployment of intelligent high-speed rail based on RIS can be designed.

### 2.1 Suppression of Doppler Effect

In the high-speed rail scenario, the train speed is much higher than that of the general terminal, so its Doppler frequency shift and expansion are more serious. Another serious problem is that when the train passes through the base station, the Doppler frequency shift will jump from +fmax to -fmax. The sudden positive and negative jump of Doppler frequency shift will make it difficult for the receiver to accurately compensate the frequency shift. Severe Doppler effect is one of the typical characteristics of high-speed rail channel propagation.

RIS' ability to dynamically control the amplitude and phase of wireless signal propagation provides a new design opportunity to solve the problem of Doppler frequency shift in high-speed rail scenarios. The use of real-time adjustable RIS can effectively reduce the rapid fluctuation of received signal strength caused by Doppler effect [6-7]. The running direction and regular running track determined by the high-speed railway wireless communication scene lead to the presentation of regular and predictable Doppler frequency shift curve, which makes it easy for the receiver and transmitter to track and compensate Doppler effect [2]. The influence of Doppler frequency shift mainly lies in the difference and dynamic change of Doppler frequency shift of different multipath signal components reaching the receiver. The main scatterers of the wireless communication channel along the high-speed rail are relatively regular and relatively determined, so RIS can be deployed on the surface of some key main scatterers, and the Doppler frequency shift alignment of different multipaths can be adjusted and controlled based on measurement and prediction, so as to reduce the impact of Doppler frequency shift. There are two special scenarios that need to be optimized: the high-speed rail is close to RIS, but the car body is on one side of RIS; The train passes through RIS, and the front and rear carriages of the car body are located on both sides of the carriage.

(1) For the scene where the high-speed railway is close to the RIS, the carriages in different parts have different angles relative to the RIS, so there are different Doppler shifts. The RIS reflecting surface can be divided into blocks, and the antenna array elements of different sub blocks independently beam the signals incident on it to align the carriages of different trains, and adjust and control to compensate for different Doppler frequency shifts.

(2) For the scene where the train passes through RIS, if RIS serves both sides of the carriage at the same time, there will be the above Doppler frequency shift positive and negative jump effect. In order to reduce the impact of this effect, an optional scheme is to use the RIS in the direction of train travel to serve the carriage in the front of the train, while the RIS in which the train is passing only serves one side of the carriage, so as to avoid the problem of Doppler frequency jump. RIS is simple, easy to deploy and low-cost, and can be deployed more intensively along the railway, so it provides the possibility for the above scheme.

### 2.2 Solve The Problem of Frequent Handover

The high-speed movement of high-speed rail above 350 km / h will cause frequent cell handover behavior, which may cause problems such as the decline of network throughput and the increase of service interruption probability, thus affecting the high-speed rail communication experience.

At present, there are two solutions to the problem of frequent handover: using radio frequency pull-out module (RRU) or distributed antenna to expand the coverage of the cell, so as to reduce the handover frequency; Optimize the inter cell handover process to minimize the performance impact of frequent handover. The first type is the more important solution, but it needs to deploy more rrus or distributed antennas. RRU or distributed antenna has high price, large volume and weight, high power consumption, and requires more broadband return links. Therefore, this scheme has great challenges in site selection, return link deployment, power supply and other aspects.

Based on its own technical characteristics, RIS has three ways to replace or improve traditional solutions:

(1) The traditional distributed antenna nodes are completely replaced by low-cost reflective RIS, and the coverage line length is extended, so as to reduce the handover frequency of the cell. RIS of reflection mode is further divided into passive reflection RIS and active reflection RIS. For passive reflection RIS, the RIS beamforming gain can be used to enhance the signal and expand the coverage. Passive reflection RIS is low cost, low power consumption, simple and easy to deploy. Only relying on the beam forming gain can not amplify the signal amplitude, so the coverage expansion range is limited. Active reflection RIS can amplify the reflected signal, so it can further expand the coverage. Active reflection RIS has higher requirements for cost, complexity, deployment, etc. Compared with the traditional distributed antenna scheme, the requirements of RIS in both reflection modes are greatly reduced.

(2) The traditional distributed antenna is combined with RIS. Considering the limited coverage of RIS, the deployment density of distributed antennas can be reduced to a certain extent, and RIS can be used to enhance the coverage.

(3) The transmission RIS is used to improve the traditional massive multiple input multiple output (MIMO) antenna, that is, the passive transmission RIS is used to replace the traditional active phased array antenna, so as to reduce the size, weight, power consumption and cost of the antenna. In addition, the passive transmission RIS is used to replace the traditional active phased array antenna, which is convenient to make some special-shaped antennas, so as to better meet the deployment requirements of high-speed rail under different natural conditions. For example, for the semi cylindrical surface shape, the beam coverage angle can better align the trains at different angles.

In addition, the demand for high-speed rail communication services will migrate as a whole with the movement of high-speed rail, that is to say, only the communities through which the high-speed rail passes need to be connected; After the high-speed railway passes through and before the next train arrives, the community does not need to support high-speed railway communication. Then, the adjacent base stations (nb_k and nb_k+1) along the railway can share the RIS between them, so as to reduce the coverage cost as much as possible. When the traditional two adjacent base stations share the remote RRU or distributed antenna, they need to switch the service data and control signaling of large bandwidth with low delay, so the implementation complexity is high; However, shared RIS only needs to switch low bandwidth control signaling between base stations, and the delay requirement of RIS control signaling can be moderately reduced, so as to achieve a balance between the dynamics of RIS control and the real-time performance of shared switching.

## 2.3 Overcoming High Penetration Loss

The penetration loss of high-speed rail includes two cases: the penetration loss of the train metal carriage and the penetration loss of the train window glass. In this section, we first analyze two types of penetration loss, and then discuss different solutions.

## 2.3.1 Penetration Loss Analysis

**(1) Metal carriage**

The metal carriage of high-speed railway will bring serious penetration loss, which makes the signal coverage in the carriage face severe challenges. The reference values of penetration loss of several typical vehicle models are given in the literature. The loss value of CRH3 is higher, which is 24-26 dB (see Table 1). In addition, for the same vehicle, different signal incident angles will also correspond to different penetration losses. When the wireless signal is vertically incident on the car, the corresponding penetration loss is the minimum; On the contrary, the smaller the incidence angle of the wireless signal, the greater the penetration loss. Therefore, the closer the base station is to the railway, the smaller the incident angle of the edge signal of the coverage area entering the carriage and the greater the penetration loss. Therefore, reasonable control of the incident angle can better meet the high-speed track coverage target.

Table 1 penetration loss of common high-speed train models and signals (frequency point: 1.8 GHz)

| Model | Train material | Loss reference value /db |
|---|---|---|
| Ordinary train | Ferruginous | 12~15 |
| CRH1 (Lombardi train) | Stainless steel | 20~24 |
| CRH2 (some bullet trains) | Hollow aluminum alloy car body | 14~16 |
| CRH3 (Beijing Tianjin Intercity) | Aluminum alloy car body | 24~26 |
| CRH5 (Alstom) | Hollow aluminum alloy car body | 22~24 |

**(2) Glass window**

The penetration loss of train window glass is lower than that of metal carriage. However, due to the special needs of anti-collision, double-layer glass of special materials is generally used, so its penetration loss can not be ignored. Especially in millimeter wave, the transmission loss is obvious. Reference [9] gives the penetration loss values of several typical glasses in the 28 GHz band, such as transparent glass (3.6 ~ 3.9 dB) and colored glass (24.5 ~ 40 dB). In addition, the area of the glass window is relatively small, and the incident energy of the signal is limited. Similar to the metal car, the penetration loss of the window glass is also affected by the incident angle of the signal. The measurement results provided in reference [10] show that the penetration loss increases with the increase of the incident angle of the signal. However, the measurement results in this paper also show that the signals in different polarization directions vary greatly with the incident angle. The transmission coefficient of vertical polarization signal decreases linearly with the increase of incident angle; The transmission coefficient of the horizontally polarized signal remains stable at different incident angles. This phenomenon gives us a new enlightenment: since signals with different polarization directions have different transmission loss changes at different incident angles, we can use this phenomenon to improve the influence of transmission loss. For example, a signal in the horizontal polarization direction insensitive to the incident angle may be considered to penetrate the window glass.

## 2.3.2 RIS Overcoming Penetration Loss

There are two main modes to realize the internal coverage of wireless mobile network in high-speed railway carriages: the signals of nodes along the railway directly penetrate the train metal and window glass into the carriages, but there are serious penetration losses; To deploy mobile nodes on the train, such as on-board relay or customer front-end equipment (CPE), it is necessary to open holes on the top of the car to deploy antennas to receive signals into the car. RIS has different solutions for the above two modes, which will be discussed separately below.

**(1) Window Glass Layout Transparent Form RIS**

The size of train windows is limited, and the beam of communication nodes along the railway (especially in the millimeter wave band) is easily blocked by metal carriages. In addition, the signal will also have large penetration loss through the window glass. On the premise of not changing the layout of high-speed railway carriages, the transparent RIS is arranged at each window, which can effectively enhance and cover the signals outside the carriages. If RIS is dynamically adjustable, it can be adjusted in real time to track users from different angles in the covered carriage, so as to solve the problem of beam blocking and obtain beam shaping gain; For the simplified fixed weight RIS, a wide beam adjustment weight can be designed to solve the beam blocking problem, but the beam forming gain cannot be obtained. Of course, RIS can also be deployed inside the car at the same time, and the RIS can control the beam coverage through transparent auxiliary windows. In this way, we can overcome the drawbacks of the above fixed weight transparent RIS. It should be noted that the RIS design in transparent form must take into account both signal enhancement and visible light transparency.

**(2) RIS Deployed Along The Railway**

Without increasing the base station density, the rational deployment of RIS along the railway can alleviate the problem of train penetration loss to a certain extent. First of all, taking advantage of the simple and easy deployment characteristics of RIS, more RIS can be deployed along the railway to enhance the signal strength of base stations and overcome large-scale road losses. Secondly, the large-scale RIS antenna array can be used to improve the beam forming gain to better align the incident windows. In addition, the penetration loss of some window glasses is sensitive to the polarization mode of the incident signal. Based on this characteristic, RIS can adjust the polarization direction of the incident signal, so as to reduce the impact of the incident angle on the penetration loss of the window glass. As described in section 2.3.1, the transmission coefficient of vertical polarization signal decreases linearly with the increase of incident angle; The transmission coefficient of the horizontally polarized signal remains stable at different incident angles. Therefore, we can fully consider using RIS deployed along the railway to control the signal polarization direction and adjust the signal incident on the window to the horizontal polarization direction, so as to reduce the window glass penetration loss at large incident angles.

**(3) Enhance Existing Roof Antenna**

Although using on-board relay or CPE to introduce / export signals into / out of the carriage is an effective means to solve the high-speed rail communication, the deployment of traditional active roof antenna is extremely challenging to the safety of train operation and the aerodynamic characteristics of the carriage shape. As a sub wavelength passive two-dimensional hypersurface, RIS is easy to be designed into a variety of shapes to enhance the existing roof antenna, which can overcome the disadvantages of the traditional active antenna. For example, RIS can be attached to the roof surface, which is easy to install without changing the surface shape of the car. In addition, the shape attached to the roof surface is also easy to deploy RIS with larger antenna aperture to provide higher antenna gain.

**(4) RIS is Deployed on The Inner Wall of The Carriage**

Further deployment of RIS on the inner wall of the car can better regulate the propagation channel in the car and solve the problem of seat blocking. For example, RIS is deployed on the inner surface of the car or with other suitable fixed decoration. These RIS can fine tune the limited intensity signals transmitted into the compartment to align the target UE, so as to make full use of the signal energy entering the compartment. Generally speaking, the terminal in the compartment moves slowly or is almost static (such as IOT terminal). Ris-ue is a slowly varying channel, which makes channel estimation and beam tracking easier.

## 2.4 Support High-precision Positioning of Trains

Train positioning technology can obtain the information of train location, real-time speed and so on anytime and anywhere, which is the intelligent basis of realizing train dispatching and control. The latest development of 5g technology, especially the specific requirements of sub meter positioning accuracy, makes wireless network-based positioning an important technical choice for the future train positioning system [11]. RIS has the characteristics of large-scale antenna array element and large antenna aperture, which can provide higher spatial resolution, so it naturally has the ability of high-precision positioning. RIS can be used for 3D image recognition with certain accuracy in high-speed rail communication scenarios with high-frequency millimeter wave band. Based on RIS, it can accurately locate the vehicle position, accurately measure the vehicle speed, and even 3D imaging. It can monitor and control the vehicle condition to ensure safety and better support intelligent driving and scheduling.

If we want to make full use of RIS super large antenna array elements and super large antenna aperture to obtain high-precision positioning, we need to deploy RIS with certain density and measurement sensing capability along the railway, that is, these RIS panels need to be equipped with active antenna arrays with sufficient density and measurement capability, so that we can accurately estimate the channel state information between the target object and RIS. Compared with RIS with only passive antenna array, RIS with active antenna array has higher complexity and cost. It is necessary to comprehensively weigh the relationship between complexity and cost and train positioning accuracy requirements during deployment.

Compared with the transceiver based on the traditional high-speed railway communication network for wireless sensing and positioning, the introduction of RIS for electromagnetic environment sensing and positioning has three main advantages: RIS is easier to be deployed in large scale along the railway, and can realize environmental sensing and positioning without blind spots along the railway; A large number of active units that make up RIS can collect rich information, so as to obtain high-precision, fine-grained electromagnetic environment perception results along the high-speed rail; The large amount of data information obtained can be applied to data-driven artificial intelligence technology, and then more comprehensive and accurate electromagnetic environment information along the high-speed rail can be mined.

## 3 Key Technologies of RIS Enabled Smart High-speed Rail Wireless Communications

RIS enabled wireless communication involves many technologies, such as RIS hardware structure and regulation, baseband algorithm, network architecture and networking, etc. in this paper, we only discuss several key technologies that need special optimization in intelligent high-speed rail communication.

### 3.1 RIS Channel Measurement And Modeling in High-speed Rail Scenarios

Most of the RIS research work under conventional scenarios is based on simple mathematical models. At present, there are only preliminary measurement and simple modeling research [12], and there is no accurate and available RIS channel model; For the special scenario of intelligent high-speed rail communication, there is a lack of necessary measurement and modeling research. The introduction of RIS will bring great challenges to the channel measurement and modeling of intelligent high-speed rail communication, mainly including the following aspects:

(1) RIS may be used in typical frequency bands and business scenarios of intelligent high-speed rail communication, so it is necessary to consider the challenges brought by the typical frequency bands of intelligent high-speed rail communication, wireless channel measurement, channel characteristics and channel models of typical scenarios.

(2) The introduction of RIS will change the channel relationship between base station (BS) and UE, and increase the bs-ris-ue cascade path and the propagation path between ris-ris. In the high-speed rail network, RIS is deployed along the railway, on the window glass of the train, on the roof of the train and inside the train compartment, etc. These deployment methods will lead to great differences in the propagation channels of RIS. For example, when RIS is deployed along the railway, the channel between nb-ris changes slowly, and the channel between ris-ue changes rapidly; When RIS is deployed on the train, the channel between nb-ris changes rapidly, and the channel between ris-ue changes slowly.

(3) Some special scenarios of deploying RIS along the railway also bring challenges to the measurement and modeling of RIS channel. For example, when RIS is deployed in tunnels, bridges along the line, and stations along the line, it is necessary to conduct special channel propagation characteristics measurement and channel modeling research for corresponding scenarios.

Accurate measurement and modeling of RIS enabled intelligent high-speed rail wireless communication channel is a new challenge, which needs to continue to increase research investment, and lay the foundation for the design, network optimization and performance evaluation of RIS based intelligent high-speed rail communication system.

### 3.2 Channel Estimation And Feedback

In conventional scenarios, the channel estimation of RIS enabled wireless networks mainly faces two challenges: the RIS channel is composed of the joint channels between nb-ris and ris-ue, and the estimation of two joint channels needs to be considered; RIS generally has a large number of antenna arrays. For the special scenario of intelligent high-speed rail communication, its channel characteristics can be used to eliminate the impact of the above challenges as much as possible, and reduce the complexity and feedback overhead of RIS channel estimation.

The intelligent high-speed rail communication channel has four main characteristics:
(1) The channel along the railway line changes regularly with the train moving forward;
(2) The angle domain of the channel is sparse, especially in the high-frequency millimeter wave band;
(3) Channel changes have distinct geographical / location correlation;
(4) The terminals on the train migrate as a whole, and the channels of UE group have common

characteristics.

In addition, in conventional deployment scenarios, nb-ris is generally a slow changing channel, while ris-ue is a fast changing channel [13]. For the smart high-speed rail communication scenario, RIS can be deployed on the window glass. At this time, nb-ris is a fast changing channel, and ris-ue is a slow changing channel.

With the high-speed railway running along the fixed track, the propagation characteristics of the intelligent high-speed railway communication channel change regularly. Therefore, in the scenario of intelligent high-speed rail communication, the network may not need complete and accurate channel state information (CSI) feedback, and the codebook feedback with limited space quantization accuracy can meet the demand of feedback accuracy. Based on the prior track and speed of the train, the forward prediction estimation can be accurately realized, so as to obtain the channel and feedback. Due to the complexity of wireless channels, the generalization performance of AI for wireless communication is limited. For the scenario of high-speed rail communication with regular changes, AI can be used to optimize the communication performance first. For example, for the codebook design of channel quantization, traditional methods such as discrete Fourier transform (DFT) can be used for quantization, and AI training can also be used to obtain the appropriate codebook set. For codebook feedback, we can use the change regularity and running speed to feedback the appropriate codebook set and its change regularity. Further, according to the relationship between the geographical location of the train and the channel state, the codebook set for the geographical location along the specific railway is designed, so that the limited number of codebooks can be used to accurately quantify the channel space. In order to improve the accuracy of codebook based feedback, it can be supplemented by feedback of certain channel calibration information to adjust and calibrate the codebook. For example, the implementation process of updating codebook set based on geographical location is: pre configure the corresponding codebook set AK based on the geographical location of base station K; When the train passes the base station, the base station configures the codebook set AK to the UE on the train; Based on the measurement, UE selects the appropriate codebook from AK and feeds it back to base station K.

In addition, the network can predict the channel state at the next time point by using the information such as speed estimation, position and position change estimation, without real-time channel estimation feedback; When the UE set on the train is migrated as a whole, the channels of UE group have common characteristics in signal direction of arrival (DOA) / angle (AOA), speed, Doppler frequency shift and so on. These common channel parameters of UE group only need to be fed back once.

These two typical characteristics can further reduce the feedback overhead and solve the problem of feedback delay.

### 3.3 Beamforming

By adjusting the phase of each unit of RIS, the beam can be adjusted to transmit signals in a specific direction, so as to reduce the transmission power of the required signals, improve spectral efficiency, expand coverage and weaken interference. In the traditional multi antenna cellular network, beamforming design is mainly to design the pre coding and decoding matrix of the multi antenna transceiver to realize signal directional transmission. In RIS aided communication system, the high-speed time-varying channel environment of intelligent high-speed rail communication makes the system beamforming design more complex. Fortunately, we can use the regular position and movement rules of intelligent high-speed rail communication to design beams to reduce the complexity of beam forming [14].

For RIS along the railway, its beamforming design includes two typical cases: far-field scenario and near-field scenario. The far-field scenario refers to the situation that the train is far away from ris/nb. This scenario has the characteristics of high road loss, high penetration loss, relatively slow angle change, and overall shift of Doppler frequency in one direction. Therefore, beamforming is needed to enhance the signal to overcome the path loss. The narrow beam mode can better align the beam with the carriages of different parts of the train by using the unique angle change law and moving speed of the train. Near field scenario refers to the situation where the train is close to ris/nb or the train passes through ris/nb. This scenario has the characteristics of low road loss, relatively low penetration loss (the angle of signal incident on the window and compartment is small), fast angle change, Doppler frequency shift jump, etc. However, the near-field scene has low demand for beam forming gain, so wide beam or broadcast mode can be considered to solve the tracking problem of rapid angle change. As described in Section 2.1, there is a special case in the near-field scenario, that is, when the train passes through RIS, the Doppler frequency shift will jump, and the beam forming direction needs to be carefully designed. One possible design is to use multiple adjacent RIS, and the beam of each RIS is aligned only with the carriage covering one side of the RIS, so as to avoid the Doppler frequency shift jump problem. According to the requirement that multiple users or cells of the train share one RIS in the intelligent high-speed rail communication, the antenna array elements of the RIS reflecting surface can be divided into blocks, and the antenna array elements of different sub blocks respectively carry out beamforming with different weights for the incident signals of different ues or different cells [15].

In order to reduce the complexity of beamforming, we can use the regularity of high-speed railway communication channel to quantize the channel space. According to the geographical location and motion law, the precoding matrix set, precoding matrix switching law and switching speed of beamforming are designed. Further, in order to avoid the spatial quantization deviation caused by the fluctuation of channel characteristics, the channel difference can be estimated semi statically, and the prior pre coding set, switching rule, switching speed and other parameter sets can be modified / updated.

In addition, the appropriate design of intelligent high-speed rail communication beamforming can also improve the cell capacity. Assuming that the deployment density of RIS is high enough, the train can be covered by multiple adjacent RIS beams at the same time. At this time, narrow beam can be used to reduce the number of train carriages covered by each RIS beam as much as possible. Because different RIS wave numbers cover different parts of the carriage of the train, it is equivalent to achieving smaller sector division through narrow beam, improving the spatial isolation, so as to improve the frequency reuse coefficient and achieve the purpose of significantly increasing the cell capacity.

### 3.4 Network Architecture And Deployment

In this section, we will comprehensively discuss RIS network architecture design and network deployment based on the characteristics of intelligent high-speed rail communication requirements and RIS technical characteristics.

Literature [16] points out that the communication requirements of intelligent high-speed rail can be divided into four categories: continuous wide area coverage of railway main line, hot areas such as railway stations and hubs, monitoring of ground infrastructure along the railway, and broadband application of intelligent trains. The demand for intelligent high-speed rail communication has distinct characteristics, which is mainly manifested in the extremely uneven distribution of capacity and coverage demand in geographical space, with distinct regularity, that is, the business demand is only limited along the railway; With the overall migration of high-speed train operation, it is reflected in the overall migration of group switching, capacity and coverage demand; Linear distribution along the railway line.

Reference [17] gives the traditional railway communication network topology based on RRU, as shown in Figure 1. For the typical deployment mode of high-speed rail communication based on RIS, based on the environment and channel characteristics of high-speed rail, and combined with the technical characteristics of RIS, literature [3] proposed three deployment modes, as shown in Figure 2. In this paper, we make a supplement and summarize four typical network deployment modes: deploying RIS along the railway; RIS is deployed on the top of the carriage for the enhanced antenna of high-speed rail mobile relay or CPE; The window glass is deployed with transparent reinforced RIS; RIS is deployed on the inner wall of the carriage. The second category is new.

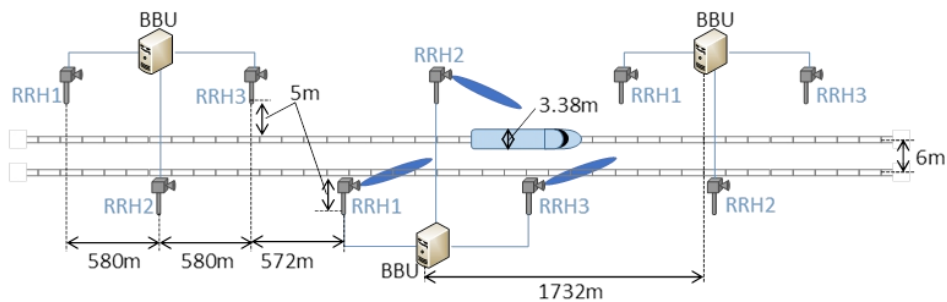

BBU: baseband processing unit
RRH: RF remote head
Figure 1 railway communication network topology based on radio frequency remote unit [17]

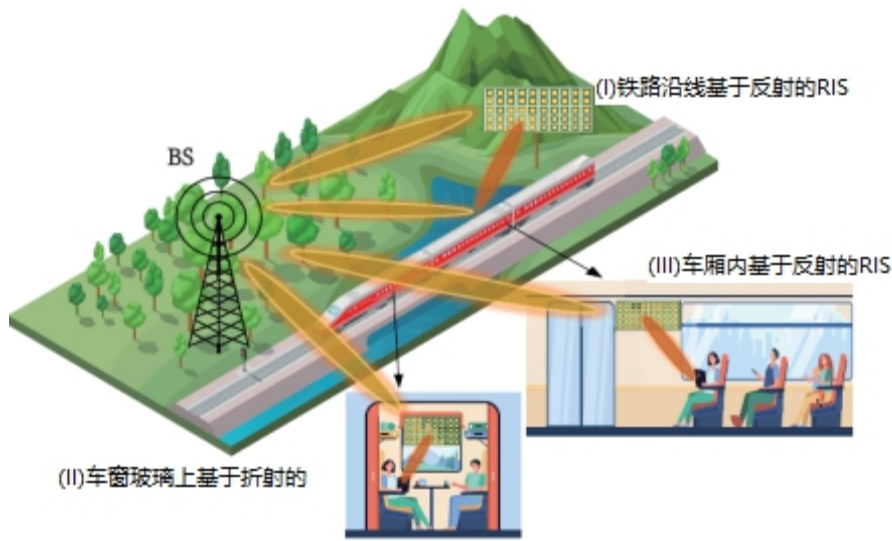

BS: base station
RIS: Reconfigurable Intelligent Surface
Figure 2 high speed rail communication scenario assisted by RIS [3]

The demand for high-speed rail communication service has the characteristics of overall migration with the movement of high-speed rail, that is, only the communities through which high-speed rail passes need service connection. After the high-speed railway passes through and before the next train arrives, the community does not need to support high-speed railway communication. A very natural idea is: as mentioned in Section 2.2 above, adjacent base stations along the railway can share RIS between them by relay. With the shared RIS scheme, only low bandwidth control signaling needs to be switched between base stations, and the delay requirements of RIS control signaling can be moderately reduced to achieve a balance between the dynamics of RIS control and the real-time performance of shared switching. However, when two adjacent base stations in the traditional high-speed rail network share RRU or distributed antenna, it is necessary to switch the service data and control signaling of large bandwidth with low delay, and the implementation complexity of this process is high.

The base station controls the return link of RIS. Different deployment methods have different design constraints, so the optional implementation methods are also different. For the RIS deployment mode along the railway, the return link between the base station and RIS can adopt wired or wireless communication mode. The connection mode of wireless return is flexible to deploy, but it needs to occupy spectrum resources to transmit the control signaling between nb-ris, so there will be some spectrum resource overhead. However, the return link control signaling information rate is low, and the overhead of spectrum occupation is not high. For the two modes in which RIS is deployed in the window glass of the train and inside the carriage, the return link between RIS and the base station can obviously only use wireless communication. RIS deployed on the roof to enhance the mobile relay or CPE antenna is controlled by the mobile relay or CPE. The return link is connected with the mobile relay or CPE and is generally connected by wire.

The above deployment modes need to ensure the air interface synchronization relationship between base stations and between base stations and RIS, so as to ensure the synchronization relationship between RIS amplitude and phase regulation and channel / signal. Especially when multiple RIS beamforming services a UE at the same time, similar to the joint transmission with traditional comp, precise time synchronization and phase alignment are required.

RIS can be used for tunnel coverage to enhance the traditional antenna form of the existing distributed antenna system (DAS), which is smaller and easy to deploy on the side of the tunnel wall without obvious protrusion. Considering the low-cost characteristics of RIS, more passive reflection RIS can be deployed at the tunnel wall side, and the regulation and enhancement of signal coverage in the tunnel can be realized through high-density RIS.

## 4 Future Research Trends and Challenges

RIS enabled future intelligent high-speed rail communication still faces many technical problems, deployment problems and challenges of standardization process. For the RIS key technologies and solutions in the special application scenarios of high-speed rail communication, people need to conduct in-depth research and comprehensive evaluation, especially in the following aspects:

(1) After the introduction of RIS, the channel measurement and modeling of intelligent high-speed rail communication;

(2) Design and optimization of channel estimation and beamforming in RIS enabled intelligent high-speed rail communication network;

(3) RIS enabled intelligent high-speed rail communication network coexists with multiple networks along the railway;

(4) Multi user and multi service type coexistence optimization under RIS enabled intelligent high-speed rail communication network;

(5) A variety of RIS form designs suitable for different scenarios of intelligent high-speed rail communication network;

(6) Research and optimization of RIS enabled intelligent high-speed rail communication network architecture and deployment.

# 5 Conclusion

In this paper, we systematically and deeply discuss the key technologies, opportunities and challenges of RIS in the application of intelligent high-speed rail communication scenarios. As a two-dimensional implementation of metamaterials, RIS has the characteristics of low cost, low complexity and easy deployment. By building an intelligent and controllable wireless environment, RIS will bring a new communication network paradigm to the future 6G to meet the needs of future mobile communications. The simplified version of RIS will probably achieve preliminary commercial deployment and standardization in the 5g / 5G adv phase, especially for improving 5g millimeter wave coverage.

As an important intelligent transportation infrastructure, intelligent high-speed rail is a comprehensive leading field in technology and industry in China, and belongs to the category of "new infrastructure". RIS is a technology independently proposed by China and triggered global follow-up. In the future, it will become one of the potential breakthroughs in China's basic original technology and the whole industry chain. The combination of the new infrastructure of intelligent high-speed rail and the new infrastructure of electromagnetism built by RIS will obtain broader technological and industrial prospects in the future.

**Bios**

**Zhao Yajun**, Chief engineer of technical pre-research of ZTE Wireless Research Institute; At present, he is mainly engaged in the research of 5G standardization technology and 6G; The main research fields include smart hypersurfaces, spectrum sharing, terahertz communication and flexible duplex; More than 200 4G/5G related invention patents have been applied, of which more than 20 have been written into the 4G/5G standard.

**Zhang Jiayi**, Professor of Beijing Jiaotong University; The research direction is large-scale MIMO and intelligent wireless communication; He was awarded the Asia Pacific Outstanding Youth Award of IEEE communication society and the first prize of natural science of China Electronics Society; He has published more than 100 papers.

**AI Bo**, deputy secretary and vice president of the school of electronic information engineering, Beijing Jiaotong University, and deputy director of the State Key Laboratory of rail transit control and safety; The research direction is high-speed rail intelligent communication; Won the first prize of natural science of the Chinese society of electronics and the outstanding young scholars at the forefront of Engineering in China; He has published more than 300 papers.